# Configurable γ Photon Spectrometer to Enable Precision Radioguided Tumor Resection

Rahul Lall, *Member, IEEE*, Youngho Seo, *Senior Member, IEEE*, Ali M. Niknejad, *Fellow, IEEE*, and Mekhail Anwar, *Member, IEEE*

*Abstract*—Surgical tumor resection aims to remove all cancer cells in the tumor margin and at centimeter-scale depths below the tissue surface. During surgery, microscopic clusters of disease are intraoperatively difficult to visualize and are often left behind, significantly increasing the risk of cancer recurrence. Radioguided surgery (RGS) has shown the ability to selectively tag cancer cells with gamma (γ) photon emitting radioisotopes to identify them, but require a mm-scale γ photon spectrometer to localize the position of these cells in the tissue margin (i.e., a function of incident γ photon energy) with high specificity. Here we present a 9.9 mm² integrated circuit (IC)-based γ spectrometer implemented in 180 nm CMOS, to enable the measurement of single γ photons and their incident energy with sub-keV energy resolution. We use small 2 × 2 μm reverse-biased diodes that have low depletion region capacitance, and therefore produce millivolt-scale voltage signals in response to the small charge generated by incident γ photons. A low-power energy spectrometry method is implemented by measuring the decay time it takes for the generated voltage signal to settle back to DC after a γ detection event, instead of measuring the voltage drop directly. This spectrometry method is implemented in three different pixel architectures that allow for configurable pixel sensitivity, energy-resolution, and energy dynamic range based on the widely heterogenous surgical and patient presentation in RGS. The spectrometer was tested with three common γ-emitting radioisotopes ($^{64}$Cu, $^{133}$Ba, $^{177}$Lu), and is able to resolve activities down to 1 μCi with sub-keV energy resolution and 1.315 MeV energy dynamic range, using 5-minute acquisitions.

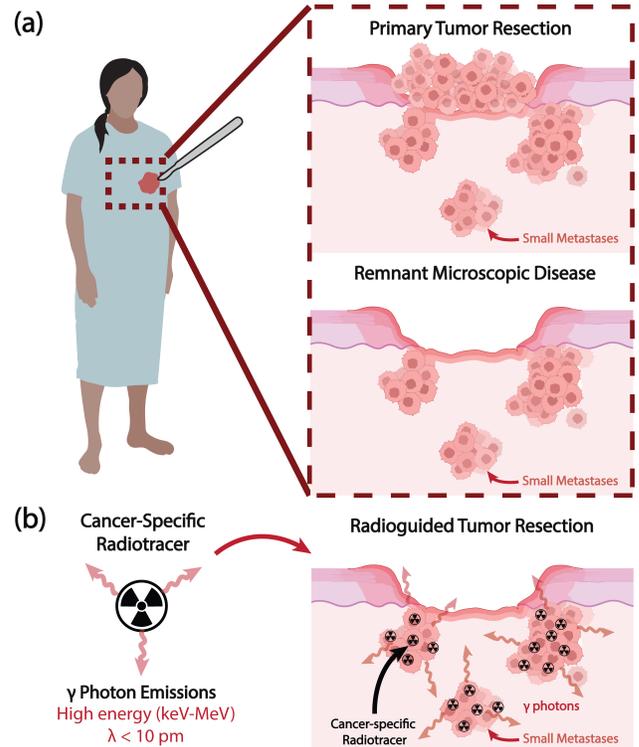

**Fig. 1:** Overview of Radioguided Tumor Resection **(a)** Current tumor resection workflow. **(b)** Use of γ-emitting radiotracer to identify and resect remnant cancer cells.

## I. INTRODUCTION

Approximately 42% of all cancer types are diagnosed at stage I or II, with 60% of these cancer patients undergoing surgery to remove a primary tumor(s) [1]-[3]. The removal of all cancer cells during surgery is vital for optimal clinical outcomes [4][5], but microscopic clusters of cancer cells in the tissue margin and at centimeter-scale depths below the tissue surface are intraoperatively hard to visualize and are often left behind, significantly increasing the risk of cancer recurrence (Fig. 1(a)). Therefore, in addition to the removal of abnormal tissue, a portion of normal tissue (> 1 cm healthy tissue margin [6][7]) is also commonly removed due to the uncertainty in assessing the microinvasion of cancer cells during surgery. There is currently a lack of real-time information regarding the presence of microscopic disease during surgical resection, with heavy reliance on these conservative positive tissue margins and confirmation of the extent of disease via pathology [6]. Pathology is typically done over the course of a week after surgery and if significant healthy, mutation-free tissue margin is not seen, the patient often undergoes surgery again to remove another conservative amount of normal tissue. Successful primary tumor(s) removal is often followed by multiple fractions of external-beam radiotherapy at the resection site or by systemic chemotherapy [8][9] to ensure that all cancer cells at/near the primary tumor

Support for this work was provided by the National Science Foundation Graduate Research Fellowship Program (DGE 2146752) and Department of Defense Idea Award. *Corresponding Authors: Rahul Lall, Mekhail Anwar.*

Rahul Lall is with the Department of Electrical Engineering and Computer Sciences, University of California, Berkeley, CA 94720 USA (e-mail: rklall@berkeley.edu). Youngho Seo is with the Department of Radiology, University of California, San Francisco, CA 94158 USA and the Department of Nuclear Engineering, University of California, Berkeley, CA 94720 USA (e-mail: youngho.seo@ucsf.edu). Ali Niknejad is with the Department of Electrical Engineering and Computer Sciences, University of California, Berkeley, CA 94720 USA. (e-mail: niknejad@berkeley.edu). Mekhail Anwar is with the Department of Radiation Oncology, University of California, San Francisco, CA 94158 USA and the Department of Electrical Engineering and Computer Sciences, University of California, Berkeley, CA 94720 USA. (e-mail: mekhail.anwar@ucsf.edu)

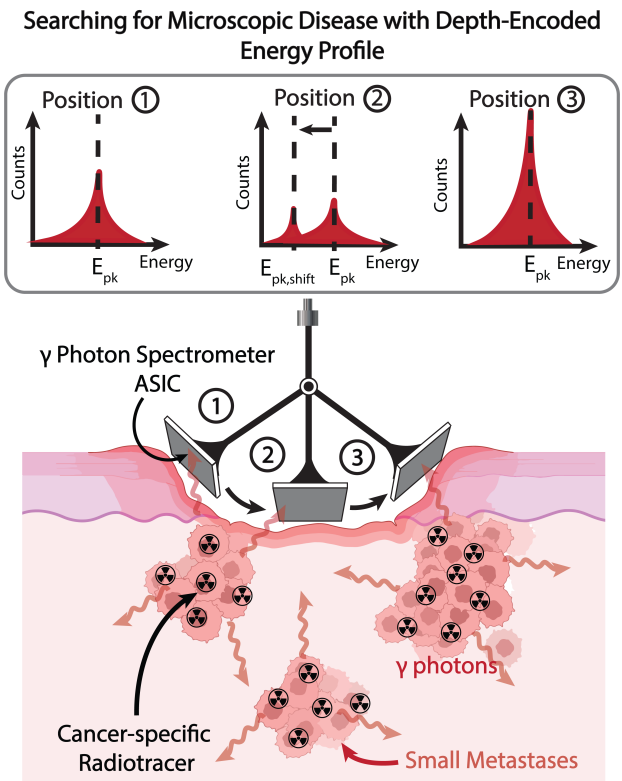

Fig. 2: Proposed γ Photon Spectrometer. A spectrometer allows for differentiation between lesion size and depth in tissue by encoding depth based on the incident γ photon energy profile.

site have been killed. Precision surgical removal of all cancer cells during primary tumor resection would reduce the amount of normal tissue removed, the number of surgical interventions, and the need for additional therapeutic interventions that may also increase secondary cancer risk (e.g., external-beam radiotherapy, chemotherapy). This necessitates a method for the real-time detection of microscopic clusters of cancer cells at/near the tumor margin.

Radioguided surgery (RGS) has shown promise in generating a signal to help achieve high confidence, precision cancer resection [10][11]. This method involves conjugating a γ-emitting radioisotope to a cancer-targeting molecule and administering this radioactive tracer intravenously to the patient before surgery. This radiotracer circulates through the blood and preferentially binds to cancer cells at or close to the primary tumor site [12]-[15]. This creates a γ photon signal that can now be used to localize these cancer cells in the tumor margin (Fig. 1(b)). During surgery, a γ photon sensor is used to locate areas with γ photon signal to remove all cancerous lesions. Since γ photons are sparsely ionizing and therefore have superior tissue penetration properties compared to visible photons used in fluorescence microscopy [16]-[18], lesions at cm-scale depths in tissue can also be sensed and removed.

To fully realize the utility of RGS, this workflow requires a γ photon spectrometer to detect both single γ photons and their incident energy. Because γ photons lose tiny amounts of energy as they traverse and scatter (for γ photon energies between 100 keV and 1 MeV) in cm-scale depths in tissue on their way to the detector, the depth of a cluster of cancer cells can be encoded by the spectrum of γ photon energies detected at the surface of the resection site (e.g., deeper lesions lead to wider spread of energies). Therefore, if incident γ photon energy can be resolved, the depth of the lesion can be as well. To illustrate this idea, Fig. 2 depicts a radiotracer with a single γ photon emission. For a shallow cluster of cancer cells with the spectrometer in position 1, there will only be a single energy peak. In position 2, the same cluster of cancer cells is present but deeper in tissue, so the signal is smaller in magnitude with some of the photons shifting to lower energies. In position 3, this larger, shallow cluster of cancer cells appears as a larger, single-energy peak. In this way, this methodology not only encodes depth but also allows for the differentiation between lesion size and lesion depth in tissue accurately to prevent excessive tissue removal. A clinically-viable γ photon spectrometer for precision tumor resection necessitates (1) a large sensing area composed of high signal-to-noise ratio (SNR) sensing elements to allow for the detection of microscopic clusters of cancer cells with low-acquisition times, (2) small form-factor and low-power consumption to allow for minimally invasive surgical intervention to reduce post-operative scarring and pain, (3) high energy resolution (i.e., sub-keV) to localize cancer cells at cm-scale depths in tissue, (4) large energy dynamic range to be compatible with many γ-emitting radioisotopes, and (5) configurability of (1)-(4) to maximize the likelihood and precision with which cancer cells can be localized even in light of surgical and patient-to-patient heterogeneity.

State-of-the-art γ spectrometers show the benefit of RGS but lack the sensing area, form factor, energy resolution, and configurability [19]-[21] to make the device clinically viable. A common approach to γ photon detection is to couple a scintillator that converts incoming γ radiation into visible or UV light, that can thereby be detected by a single photon avalanche photodiode (SPAD)-sensing frontend. These types of indirect detectors increase the probability of detection through a high quantum efficiency conversion to visible or UV light but oftentimes lose energy information in the process due to the inability to differentiate signal magnitude and incoming γ photon energy [22]-[26]. This γ photon counting approach has been explored in clinical RGS [27][28], but cannot resolve signals as a function of depth due to their inability to resolve energy. Variants of this approach that use multi-photon avalanche photodiodes (APD) can correlate single-γ photon events with detected visible light signal magnitude by measuring these events at a high frequency, but still require large voltages for biasing, present with high power consumption, have to be clocked and reset in the hundreds of MHz range, and present with scalability issues due to 3D system integration of cm-scale scintillators with the developed sensing circuitry, making this approach incompatible with RGS [29][30]. There has also been a large amount of prior work on γ photon sensing ASICs and spectrometers in the fields of high-energy physics [31]-[33], basic energy science [34]-[37], and medical physics [38]-[40]. The ASIC systems developed for these applications resolve incident γ photon energy with high energy resolution and sensitivity, but the detection scheme still involves cm-scale, energy-sensitive materials that are not inherently integrated with high-speed readout circuitry and digital systems, requiring 3D integration to an additional readout ASIC. Over the past few decades additional energy-sensitive materials have been identified to distinguish incident γ photon energies, including cadmium zinc telluride [41], perovskite [42], epitaxial silicon carbide [43], and polycrystalline diamond [44], but these non-silicon-based detection methods are all still in the cm-scale form factor and require 3D integration to a readout ASIC. These indirect approaches, therefore, complicate the mm-scale form-factors, cost, and low-power necessary for RGS implementation. Pixel-specific energy-binning approaches have been implemented in CMOS as well [45]-[47] and have shown promise in binning ionizing radiation energies in mm-scale form factors. These approaches to energy-binning suffer from low area (i.e., single pixel) per energy bin which significantly increases the acquisition time necessary to acquire enough γ photon counts to discern the presence of a cancer cell, and low number of energy bins due to the finite number of pixels. These approaches also have low degrees of configurability in regards to flux, energy resolution, and energy dynamic range, complicating their performance in face of the highly varying surgical presentation in RGS.

Therefore, although previous works [27]-[30], [31]-[40], [45]-[47] have shown promise in achieving energy-binning of γ radiation, they have few energy bins, low detection area per energy bin, no configurability, and require 3D integration with cm-scale materials making them incompatible with low-flux, wide-energy, area-constrained γ-sensing applications such as RGS. To address this gap in the state-of-art, this paper presents a mm-scale application specific integrated circuit (ASIC)-based γ photon spectrometer in a standard 180 nm CMOS process with a large number of energy bins, large sensing area per energy bin, and configurability of sensitivity, energy resolution, and energy dynamic range to encode the location of remnant cancer cells at cm-scale depths in tissue with low acquisition times [48]. The ASIC is capable of direct detection of γ photons (i.e., no 3D integration necessary) with integrated asynchronous readout for low digital readout complexity. We have characterized the spectrometer's ability to measure radioactive activity and incident γ photon energy with three commonly used radioisotopes ($^{64}$Cu, $^{133}$Ba, $^{177}$Lu) that have highly varying γ photon emissions (511 keV, 1.346 MeV; 31 keV, 81 keV, 302 keV, 356 keV; 113 keV, 208 keV).

This paper is organized as follows. Section II discusses the γ photon spectrometry method and pixel architecture used to detect single γ photons and infer their incident energy. Section III discusses the pixel architectures. Section IV discusses the spectrometer system architecture, including the configurable pixel array, SRAM sensitivity calibration unit, and decay-time processing unit. Section IV presents pixel characterization results and experimental measurements of multiple γ-emitting radioisotopes. Section V compares this system to the state-of-the-art and outlines the utility of this system for radioguided tumor resection.

## II. γ Photon Detection Architecture

This section discusses how γ photons in the energy range emitted by most radiotracers used for RGS interact with silicon. We then discuss the developed γ spectrometry method to enable low power, large sensing area energy-binning with configurable resolution and dynamic range.

### A. Compton Scattering Overview

Compton scattering is the dominant interaction mechanism for γ photon energies between 100 keV and 1 MeV in both tissue and silicon, and involves a photon scattering with a silicon orbital electron and transferring part of its original energy to that electron, with the amount of energy transferred depending on how much the photon scatters (i.e., greater scattering angle indicates larger deposited energy). RGS relies on using a radioisotope that has at least one γ emission in this energy range. The average amount of energy deposited by a γ photon that Compton scatters as a function of depth is described by the incident energy, scattering angle, and probability of interaction as seen below in Eqn. 1 [49].

$$E_{dep}|_{\Delta x} = \frac{E_{incident}(1-cos\theta)}{\frac{m_ec^2}{E_{incident}}+1-cos\theta}\left(1-e^{-\mu_{CS}|_{E_{incident}}\Delta x}\right) \quad (1)$$

where $E_{dep}|_{\Delta x}$ is the energy deposited by a γ photon in a certain depth $\Delta x$, $\Delta x$ is the distance traversed by the γ photon, $E_{incident}$ is the incident γ photon energy, $\mu_{CS}|_{E_{incident}}$ is the Compton scattering cross-section in tissue or silicon at the incident γ photon energy [48], $\theta$ is the scattering angle, $m_e$ is the mass of an electron, and $c$ is the speed of light. Compton scattering introduces two major sources of uncertainty: (1) the interaction probability (i.e., low $\mu_{CS}|_{E_{incident}}$) and energy deposited ($E_{dep}$) per interaction are both low, making single γ photon detection difficult to realize in silicon, and (2) an incident γ photon deposits a spectrum of energies in silicon depending on the scattering angle with the detector, complicating the mapping between $E_{dep}$ and $E_{incident}$.

### B. γ-Spectrometry Method

In order to enable energy resolution using direct γ detection, single γ photons have to be sensed. This is difficult to realize in this energy range due to the low probability of interaction and low energy deposition per interaction (i.e., oftentimes depositing < 1 keV). We accomplish single-γ photon detection by using very small 2 × 2 μm reverse-biased deep N-well/P-substrate diodes in a standard 180 nm CMOS process. The number of electron-hole pairs (EHPs) that are generated when a γ photon is incident on the depletion region of a reverse-biased diode is given by Eqn. 2:

$$N_{EHPs} = \frac{\frac{E_{dep}|_{x_{dep}}qf}{cos\theta_i}}{E_g} \quad (2)$$

where $N_{EHPs}$ are the number of electron-hole pairs produced per γ photon interaction, $x_{dep}$ is the thickness of the depletion region, $qf$ is the quenching factor of silicon (i.e., 1/3), $E_g$ is the bandgap energy of silicon (i.e., 1.12 eV/EHP), and $\theta_i$ is the incident angle of the γ photon onto the diode. The voltage at the diode node in response to a single γ photon is then given by Eqn. 3:

$$V_d = \frac{Q_{gen}}{C_{diode}} = \frac{qE_{dep}|_{x_{dep}} qf}{C_{diode} E_g} \quad (3)$$

where $V_d$ is the voltage signal produced at the diode-sensing node at time *t*=0, $Q_{gen}$ is the charge generated in the depletion region of the diode, $C_{diode}$ is the total parasitic capacitance at the diode-sensing node, and $q$ is the charge of an electron. The parasitic capacitance (i.e., size) of the diodes is optimized as a proxy for optimizing both the depletion width of the diode (e.g., lower capacitance, larger depletion width) and the capacitive coupling to nearby circuitry. Deep N-well diodes provide a suitable depletion width while also significantly decreasing capacitive coupling to other parts of a densely packed analog pixel (i.e., reducing the overall $C_{diode}$). From Eqn. 3, it is evident that the same charge generation at a circuit node with lower parasitic capacitance, will result in a larger voltage drop at that node. Similarly, interactions depositing more energy (i.e., larger $E_{dep}$) in the depletion region of the reverse-biased diode generate more EHPs and therefore generate a larger voltage signal at the diode-sensing node as well (Fig. 3(a)).

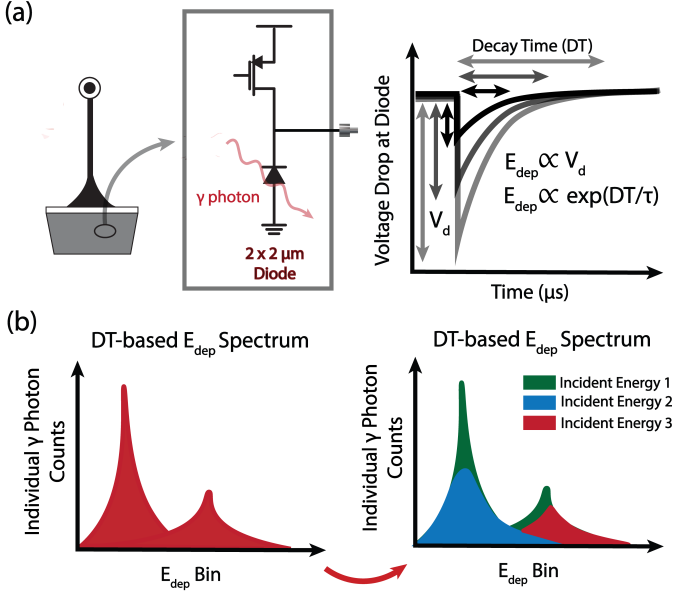

Fig. 3: Low Power and Area γ Sensing and Spectrometry Method. **(a)** Single γ photons are sensed using small reverse-biased diodes with small parasitic capacitance. To enable spectrometry, the decay time it takes for the resulting voltage signal to settle back to DC is measured in the time domain. **(b)** The *DT* measurement can be converted to the energy deposition in the depletion region which is a non-linear function of incident energy, and is therefore converted to incident energy using TOPAS, the state-of-the-art radiation simulation software.

Smaller $C_{diode}$ allows for high-SNR γ photon detection ($SNR_{V_d}$) and energy resolution ($SNR_{DT}$) as discussed below.

The PMOS device biasing the diode is biased in triode and therefore contributes a low frequency resistance of $r_{ds}$. The total integrated voltage noise power at the diode-sensing node, assuming the transistor operates above threshold, can be approximated as:

$$\bar{v}_{diode}^2 = \bar{v}_{diode,shot}^2 + \bar{v}_{res}^2 = \frac{qI_s r_{ds}}{2C_{diode}} + \frac{kT}{C_{diode}} \quad (4)$$

where $\bar{v}_{diode,shot}^2$ is the shot noise of the diode, $\bar{v}_{res}^2$ is the thermal noise from the PMOS output resistance, $q$ is the charge of an electron, $I_S$ is the reverse saturation current of the reverse-biased diode, and $r_{ds}$ is the finite output resistance of the PMOS used for biasing. The total integrated voltage noise at the diode node is band-limited in frequency by the $\tau = r_{ds}C_{diode}$ time constant at the diode-sensing node, and flicker noise is neglected in this calculation because a long-channel device is used to bias the diode. The SNR of $V_d$ at the diode-sensing node is then given by:

$$SNR_{V_d} = \frac{V_{diode}}{\sqrt{\bar{v}_{diode}^2}} = \frac{\frac{Q_{dep}}{C_{diode}}}{\sqrt{\frac{qI_s r_{ds}+2kT}{2C_{diode}}}} \propto \sqrt{\frac{2}{C_{diode}}} \quad (5)$$

As the diode size decreases, the capacitance at the diode-sensing node drops and the $SNR_{V_d}$ improves by $\sqrt{\frac{2}{C_{diode}}}$. Because there is a resistance and capacitance associated with this node, after the immediate voltage drop, the signal exponentially decays back to DC and, therefore, interactions depositing more energy also have larger decay time (*DT*). The relationship between $V_d$, $E_{dep}$, and *DT* is described by Eqn. 6 and 7 below:

$$V_{diode}(t) = V_d e^{-\frac{t}{r_{ds}C_{par}}} = V_d e^{-\frac{t}{\tau}} \quad (6)$$

$$V_d = V_{th,detect} e^{\frac{DT}{\tau}} \propto E_{dep} \quad (7)$$

where $V_{diode}(t)$ is the voltage signal at the diode-sensing node after time *t*=0, $V_d$ is the voltage signal produced at the diode-sensing node at time *t*=0, and $V_{th,detect}$ is the minimum detectable signal at the diode-sensing node (i.e., approximately 5 mV) and is the value of $V_{diode}(t)$ when *t*=*DT*. To gain insight into the $E_{dep}$ of any γ detection event, the *DT* can be measured in time (i.e., time-to-digital conversion) rather than measuring the voltage drop at the diode-sensing node directly (Fig. 3(a)). This prevents the use of high-resolution, high-power, large-area ADCs while providing an inherent logarithmic compression method for wide energy dynamic range sensing. The *DT* is given by:

$$DT = \tau \ln\left(\frac{V_d}{V_{th,detect}}\right) \quad (8)$$

The approximate total integrated *DT* noise power generated at the diode-sensing node ($\overline{DT^2}$) using the delta method [50] is given by:

$$\overline{DT^2} = \frac{\tau^2 \mathbf{V}[V_d]}{\mathbf{E}[V_d]^2} = \frac{\tau^2\left(\frac{qI_S r_{ds}}{2C_{diode}} + \frac{kT}{C_{diode}}\right)}{\left(\frac{Q_{gen}}{C_{diode}}\right)^2} \approx \frac{\tau^2}{Q_{gen}^2}(kTC_{diode}) \quad (9)$$

where $\mathbf{V}[V_d]$ is the total integrated voltage noise at the diode-sensing node and $\mathbf{E}[V_d]$ is the expected value of the voltage drop at the diode sensing node. The SNR in $DT$ is given by:

$$SNR_{DT} = \frac{DT}{\sqrt{\overline{DT^2}}} = \frac{\tau ln\left(\frac{Q_{gen}/C_{diode}}{V_{th,detect}}\right)}{\sqrt{\frac{\tau^2}{Q_{gen}^2}(kTC_{diode})}} \propto ln\left(\frac{Q_{gen}/C_{diode}}{V_{th,detect}}\right)\frac{Q_{gen}}{\sqrt{C_{diode}}} \quad (10)$$

Therefore, $SNR_{DT}$ is also maximized for a minimum $C_{diode}$. The diode size was chosen to be small but not minimum-sized to optimize both the available sensing area as well as the $SNR_{V_d}$ and $SNR_{DT}$ simultaneously.

The spectrometer can be used to measure the tumor margin for a period of time, with the output of the system being a distribution of these $DT$ measurements from all of the γ photon detection events in that period of time. These individual $DT$ measurements can then be converted to a quantity proportional to $E_{dep}$ using Eqn. 7 (Fig. 3(b)). Because the predominant interaction method for γ photons in this energy range is Compton scattering, $E_{dep}$ is a nonlinear function of $E_{incident}$, the quantity that encodes cancer cell depth in tissue. $E_{dep}$ is converted to $E_{incident}$ using the state-of-the-art radiation simulation software, TOPAS (GEANT4 Wrapper) [51]. In practice, TOPAS would be used to simulate a point source at varying depths in tissue away from the detector. At each point-source location, the $E_{incident}$ γ photon that generated a certain $E_{dep}$ in the depletion region of the diode can be tabulated. All of the $E_{dep}$-$E_{incident}$ key-value pairs can then be placed in a look-up table. The measured $E_{dep}$ distribution can be assigned to the most similar (e.g., sum of square difference) simulated $E_{dep}$ distribution in the look-up table. Using the simulated TOPAS mapping (Fig. 3(b)), the $E_{incident}$ of each $E_{dep}$ measurement can then be assigned by labeling the proportion of each $E_{dep}$ bin that was deposited by a γ photon with a certain $E_{incident}$.

The closest simulated TOPAS measurement is used as a ground truth mapping between measured $E_{dep}$ and the inferred $E_{incident}$ distribution, so this method does not require an inversion algorithm. Although more advanced reconstruction methods are outside the scope of this paper, future work will focus on different algorithms that can improve the performance of this IC for application to RGS. $E_{dep}$ events are labelled with a known set of possible $E_{incident}$ energies from TOPAS, with the uncertainty in $E_{dep}$ propagating as mislabeled $E_{incident}$ events (i.e., uncertainty). The total number of mislabeled $E_{incident}$ events is therefore given by the total mismatch between the simulated and measured $E_{dep}$ distributions. The energy resolution is therefore the full width at half maximum in $E_{dep}$, and this can be conservatively estimated as the ±2σ width.

$$\varepsilon_{incident} \propto \Delta E_{dep} < \alpha V_{th,detect} e^{\frac{DT}{\tau}}\left[e^{\frac{2\overline{DT}}{\tau}} - e^{\frac{-2\overline{DT}}{\tau}}\right] \quad (11)$$

where $\varepsilon_{incident}$ is the average number of photons with mislabeled incident energies, $\Delta E_{dep}$ is the ±2σ deposited energy resolution of the system, $\alpha$ is linear scaling factor relating the absolute $E_{dep}$ and $V_d$, and $\overline{DT}$ is the RMS noise in $DT$ from Eqn. 9. The absolute $\varepsilon_{incident}$ of the system is based on the chosen bin width ($\Delta E_{bin}$) used to bin the measured $E_{dep}$ histogram. Therefore, $\Delta E_{dep}$ is scaled by the probability of finding a count that deposits $E_{dep}$ outside of $E_{dep} \pm \Delta E_{bin}/2$. $\Delta E_{bin}$ is a parameter depending on the amount of $\varepsilon_{incident}$ error that is acceptable in a certain measurement environment.

## III. PIXEL ARCHITECTURES

Since small 2 × 2 μm diodes are used for single γ photon detection, a pixelated-sensing architecture was implemented to ensure sufficient detection area to minimize acquisition time. The $DT$ measurement, readout, and quantification circuitry used in these pixels builds on the time-over-threshold (ToT) method for pulse-amplitude measurement used in prior CMOS radiation detectors [52][53], but enables direct single γ photon detection with asynchronous readout (i.e., no integration, reset, or readout period). The pixel architectures introduced in this section allow for increased sensitivity, less digital readout complexity, and lower power consumption and area, suitable for application to RGS. Because RGS requires scattered photon detection for cancer cell depth-encoding, small acquisition times for rapid scanning of the resection site, and optimal energy resolution and dynamic range for the current measurement environment, this system has three diode-based pixel architectures that optimize for fine energy-resolution (Fig. 4(a)), low-flux applications (Fig. 4(b)), and energy-calibration (Fig. 4(c)), respectively. The details of these pixel architectures and their integration with the digital readout is described in more detail in the subsequent sections, but are outlined briefly here. Energy-resolving pixels bin the deposited γ photon energy using the proposed $DT$-based γ-photon spectrometry method. The $DT$ is estimated using a 10-bit counter whose timing is set by a configurable clock, and the period of this clock is set by the energy-calibrating pixels. The energy-calibrating pixels find the maximum $E_{dep}$ by any γ photon and assigns the minimum period to the configurable clock that supports this dynamic range (i.e., dynamic range of counter). The low-flux pixels do not add additional energy-sensing capabilities but have 3x higher fill factor with the same diode sensitivity to allow for higher count rates in low-flux environments when enabled.

Each of these pixel architectures are constrained by the same general sensing area, size, and power specifications. Sensing area was optimized by trading off sensitivity and pixel fill-factor, which are inversely proportional in this design (Eqn. 3). A minimum detectable $N_{EHPs}$ of 50 EHPs in the diode depletion region was targeted. The overall pixel size was constrained to include all the relevant amplification and custom pixel sensitivity circuitry, leading to a maximum diode size of 2 × 2 μm (i.e., $C_{diode}$ of 1.5 fF including all parasitic capacitances connected to the diode-sensing node) while maintaining high sensitivity (i.e., $V_{th,detect}$ of 5 mV). To ensure that γ photons interacting with the reverse-biased diodes are the only signals that are being amplified and identified as γ photon detection events, the capacitive loading at every other node in the pixel analog readout circuitry is set to be at least 10 times greater than at the diode-sensing node. The analog sensing and amplification does not use amplifiers with very large open-loop gains (e.g., maximum of 100) such that close to minimum device sizes can be utilized to reduce pixel size while maintaining low static errors in closed-loop. Pixels share common row lines such that very little logic or area is necessary to relay data to the digital blocks. To reduce power grid routing complexity on-chip and high-current cables during surgery, sub-1 mW operation for the whole configurable pixel array (i.e., < 0.130 μW/pixel) was targeted. The common circuit blocks used in all of the pixel architectures discussed below are shown in Fig. 4(d).

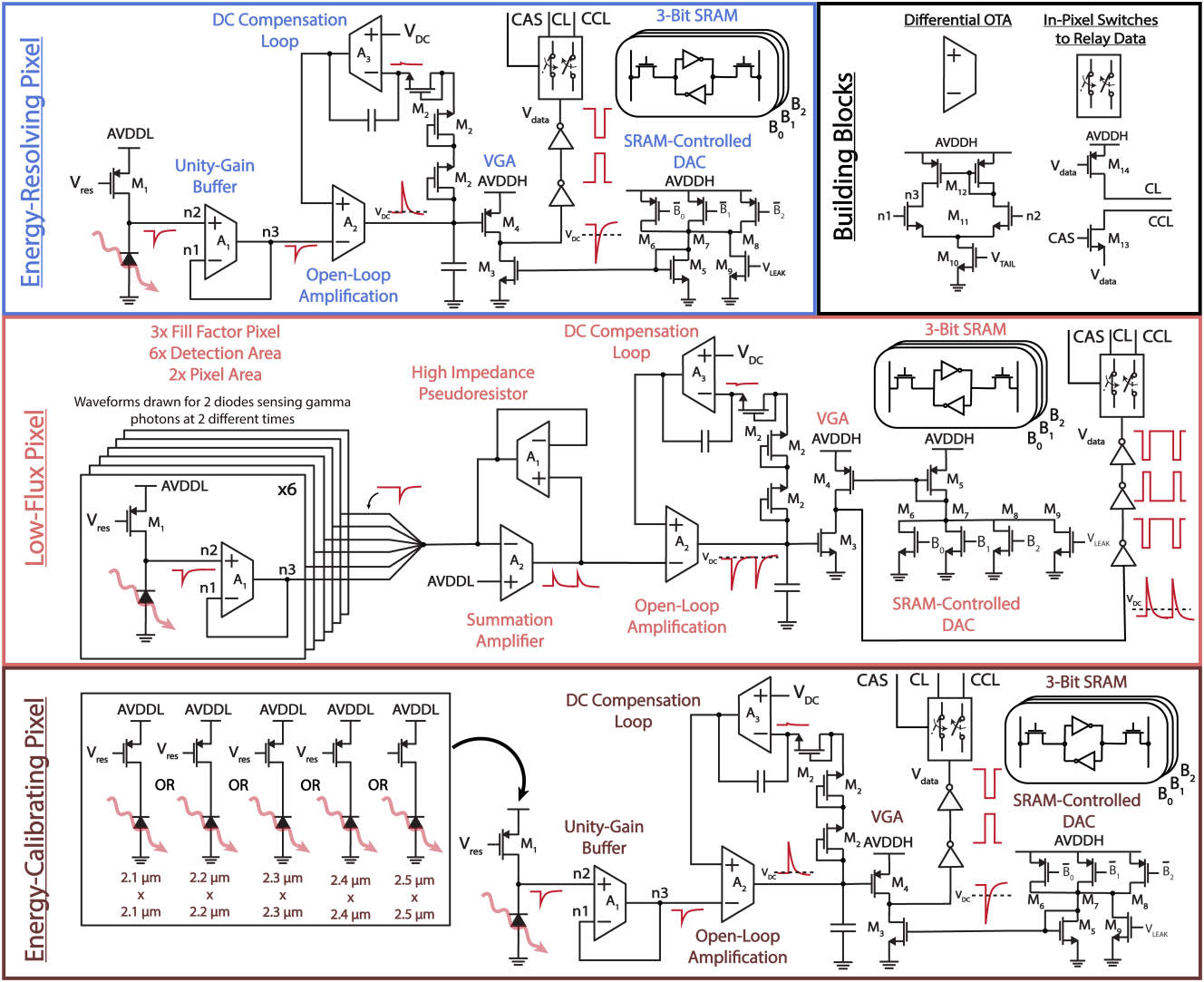

**Fig. 4:** γ Spectrometer Pixel Architectures. **(a)** Energy-resolving pixels are always enabled, and amplify and one-bit digitize the voltage signal generated from an incoming γ photon signal. The widths of these pulses are subsequently measured at the system-level. **(b)** Low-flux pixels are enabled in low γ photon flux environments and have 3x the fill factor in 2x the area compared to the energy-resolving pixels. **(c)** Energy-calibrating pixels use a gradient of diode sizes to optimize the energy resolution and energy dynamic range of the system. **(d)** Differential OTA and switch topology used in every pixel type.

### A. Energy-Resolving Pixels

Energy-resolving pixels enable large sensing area energy-binning of incoming γ photons using the energy spectrometry method discussed in the previous section (Fig. 3). Energy-resolving pixels' sensor front-end begin with a single reverse-biased 2 × 2 μm diode (Fig. 4(a)). When a γ photon interacts with the depletion region of this diode, there is a small drop in voltage at the diode-sensing node with some time constant τ, set by the $r_{ds}$ of the PMOS biasing device (M$_1$) and $C_{diode}$. To change the τ time constant at that node, the value of M$_1$'s $r_{ds}$ that reverse-biases the diode can be adjusted by modifying the gate bias voltage ($V_{res}$) that is generated by an off-chip digital-to-analog converter (DAC). Eqn. 8 expresses DT as a function of $r_{ds}$ but can be further expanded to express DT as a function of $V_{res}$:

$$DT = \frac{C_{diode}}{\mu_p C_{ox} \frac{W}{L}(V_{DDL}-V_{res}-|V_{th,p}|)} \ln\left(\frac{V_d}{V_{th,detect}}\right) \quad (12)$$

where $\mu_p$ is the hole mobility, $C_{ox}$ is the oxide capacitance of the PMOS biasing device, $V_{DDL}$ is the source voltage of the PMOS, and $V_{th,p}$ is the PMOS threshold voltage. The ability to change the time-constant at the diode-sensing node allows for the tradeoff between energy resolution and energy dynamic range (i.e., larger $\tau$ leads to better energy resolution but lower energy dynamic range, and vice versa). To reduce diode capacitive loading and maximize γ photon sensitivity, a unity-gain OTA $A_1$ with a minimum sized input-pair (parasitic capacitance of 0.35 fF) is utilized in every pixel to buffer out this voltage signal. The OTA is placed in unity-gain configuration to reduce the effect of input device mismatch on output voltage offset and to reduce the total-integrated output voltage noise power of the amplifier this early in the signal amplification chain. If an open-loop OTA is used instead, large input devices would need to be used to decrease device mismatch across the IC, reducing the sensitivity of the pixel due to the extra capacitance. The open-loop gain ($A_{OL}$) of $A_1$ was designed to be above 100 to achieve a static error of < 1% while reducing pixel area.

The output of this buffer is connected to an open-loop OTA $A_2$ with an $A_{OL}$ of approximately 25. The diode is connected to a unity-gain buffer instead of this open-loop OTA directly because of the large asymmetrical Miller amplified gate-to-drain capacitance connected from the inverting input of $A_2$ to its output. A DC compensation loop composed of an RC OTA integrator $A_3$ connects the output of $A_2$ back to its non-inverting input to mitigate any output referred voltage offset from input-pair mismatch and to secure the output DC level of this OTA to $V_{DC}$, a voltage provided by an off-chip DAC. Because this is a low-frequency feedback loop, the RC low-pass filter introduced by the three diode-connected NMOS devices and metal-insulator-metal (MIM) capacitor needs to be placed at a low enough frequency to ensure complete attenuation of the small-signal γ photon pulse that is present at the output of $A_2$. Modifying $\tau$ using Eqn. 12 changes the frequency composition of the small-signal pulse and helps ease the constraints on this RC low-pass filter. This allows for the use of a smaller MIM capacitor and size of diode-connected devices to minimize pixel area while still maintaining sufficient attenuation. $V_{DC}$ and the resulting output bias voltage of $A_2$ is set to mid-rail (i.e., 600 mV). The output of $A_2$ is routed to a variable gain amplifier (VGA). Reliably setting the input-bias of the VGA allows the VGA to modify the waveform minimally to maximize sensitivity and minimize the variance in sensitivity across all the pixels. The VGA is composed of a PMOS-input (M4) common-source (CS) stage with configurable load (M3). The load is configured by changing the gate-to-source voltage of the NMOS active load by writing to 3-bit in-pixel 6T SRAM. The 3-bit in-pixel 6T SRAM controls in-pixel DACs by changing the current that is sunk (M6, M7, M8) through a diode-connected NMOS device (M5). The DAC has a global leakage bit (M9) that can be set off-chip to mitigate any systemic offset errors as well. This VGA is used to set the custom sensitivity of each pixel to maximize the sensing area that can be utilized for measurements. The VGA also shifts the DC level at its input to its output based on the output resistance of the active load. The amplified and level-shifted pulse from the VGA is digitized using two inverters and sent to the digital DT processing unit via a set of switches that relay the data on shared row lines. Therefore, if the amplified and level-shifted pulse at the output of the VGA crosses the switching point of the inverter, a γ photon detection event is detected and a rail-to-rail voltage pulse is created. We use a VGA and inverter approach instead of the typical comparator approach to pulse generation [52][53] because this design uses very small $C_{diode}$ to enable single photon detection which may vary significantly across the chip. The VGA allows for the user to calibrate for the pixel-to-pixel variation caused by $C_{diode}$ variation by changing the amplitude and moving the DC level of the signal at the input of the inverter. This is essentially changing the switching point of the inverter to allow for a homogenous response from all pixels given the same input source.

When there is a γ photon detection event, the output of the inverter ($V_{data}$) goes low, and a PMOS pulls the count-line (CL) high. The width of these digital pulses on the CL are measured in the $DT$ processing unit using a configurable period digital clock to estimate the $DT$ of a given pulse. When $V_{data}$ goes low and the column-address sweeping (CAS) signal goes high, the count-column line (CCL) goes low. The CCL is used to help infer which column a detection event on a specific row originated from.

The overall pixel size is 26 μm × 18 μm, with a fill-factor of approximately 1/115. Energy-resolving pixels are always enabled, and the $DT$ of their pulses on the CL are quantified with the optimized clock period set by the energy-calibrating pixels.

### B. Low-Flux Pixels

Low-flux pixels enable low acquisition times even when the γ photon flux is low and can be enabled when the recorded γ photon count rates are less than a global threshold that is set off-chip (Fig. 4(b)). Low-flux pixels start with the same building block of a 2 μm × 2 μm reverse-biased diode and unity-gain buffer $A_1$ pair but utilize 6 of these pairs in 2x the area of the energy-resolving pixels, increasing the pixel fill-factor by 3x while maintaining the same diode sensitivity (i.e., minimum $C_{diode}$ by using a unity-gain OTA in the first stage). Each of the outputs of these unity-gain OTAs are tied together to add their output currents together. To convert this summed current into a voltage, an OTA $A_2$ with a feedback unity-gain OTA $A_1$ is used. The output resistance of this feedback OTA is designed to be matched to the output impedance of each of the individual OTA buffers connected to the diodes. This forms a summation amplifier circuit with a transimpedance gain of -1, where the feedback unity-gain OTA acts as a large-valued, low-area pseudoresistor. Because the total integrated output current noise power is added at the summation node of the transimpedance amplifier, the SNR at this node is lower than the energy-resolving pixels, decreasing the benefit of improved fill-factor below 3x. The total integrated thermal noise at the summation node ($\bar{v}_{sum}^2$) is approximately:

$$\bar{v}_{sum}^2 = 6 \times \frac{\frac{R_{out}^2}{4R_{out}C_{out}}\sum_{i=1}^{4}4kT\gamma_{noise}g_{m,i}}{g_{m,1}^2 R_{out}^2} \approx \frac{24\gamma_{noise}kT}{C_{out}A_{OL}} \quad (13)$$

where $k$ is the Boltzmann constant, $T$ is the temperature, $\gamma_{noise}$ is the technology white noise factor, $g_{m,1}$ is the transconductance of the input pair devices, $R_{out}$ is the output resistance of $A_1$, and $C_{out}$ is the output capacitance of $A_1$. The large $A_{OL}$ of $A_1$ ($A_{OL}$=100) minimizes the reduction in the SNR while still optimizing for pixel area.

The summation of individual γ detection events from any of the 6 diodes is routed to an open-loop OTA ($A_{OL}$=25) with a DC compensation loop composed of an RC integrator with 3 NMOS diode-connected devices (M$_2$) and a 50 fF MIM feedback capacitor, similar to the energy-resolving pixels. The sign of the voltage signal in the low-flux pixels is inverted compared to the energy-resolving pixels at this stage in the amplification chain due to the transimpedance amplifier. Therefore, a VGA with an NMOS-input (M$_3$) CS-stage with SRAM-configurable load (M$_4$) is utilized. The SRAM-configurable DACs (M$_6$, M$_7$, M$_8$) in this pixel architecture change the current supplied by a diode-connected PMOS device (M$_5$) to set the gate-voltage of the PMOS active load in the CS-stage. The DAC still utilizes a global leakage bit (M$_9$) that can be set off-chip to mitigate any systemic offset errors, similar to the energy-resolving pixels. The amplified voltage pulse at the output of the VGA is then sent through three inverters for 1-bit digitization, with the extra inverter fixing the sign of $V_{data}$ such that it is active-low whenever there is a γ photon detection event. The count data is relayed to the CL and CCL in the same way as the energy-resolving pixels.

The overall pixel size is 26 μm × 36 μm, with a fill-factor of approximately 1/38. Low-flux pixels are only enabled in low-flux environments since the current implementation of this IC does not discriminate between counts on any of the 6 diodes in the same pixel. Therefore at high-flux rates, coincident events can skew count statistics in these pixels.

*C. Energy-Calibrating Pixels*

Energy-calibrating pixels allow for the calibration of optimal energy resolution and energy dynamic range given the current measurement environment. This ensures that the maximum energy resolution is used without losing data that exceeds the energy dynamic range of the system. Their signal amplification chain is the same as the energy-resolving pixels' but with varying diode sizes (4.41, 4.84, 5.29, 5.76, or 6.25 μm$^2$) per row in the pixel array (Fig. 4(c)). Increasing the diode size, increases $C_{diode}$, decreasing $V_d$ proportionally for the same $Q_{gen}$. Larger diode sizes with the same circuitry as the energy-resolving pixels, therefore, degrade the pixel's sensitivity to γ photons. This enables coarse energy-binning to find the maximum $E_{dep}$ by a γ-emission by providing a gradient of varying sensitivity pixels. This method measures the maximum $E_{dep}$ directly at the earliest point in the analog signal chain, making it less prone to PVT variation that may be introduced by the subsequent amplifiers in the analog pixel. Based on the maximum diode size that can still detect γ photons, the period of the configurable clock used for $DT$ estimation by the $DT$ processing unit in the time domain is chosen to optimally set the energy resolution and dynamic range of the system. Once the $DT$ counting clock is set for the energy-calibrating pixels, this setting is used for the rest of the pixel array, and the sensitivity of each pixel in the array is subsequently calibrated (i.e., via in-pixel SRAM).

The overall pixel size for all energy-calibrating pixels is 26 μm × 18 μm, with fill-factors between 1/115 and 2/115. Energy-calibrating pixels can be used for measurements as long as the $E_{dep}$ is adjusted to account for the larger $C_{diode}$, but these pixels are primarily for system energy-sensing calibration.

IV. SYSTEM ARCHITECTURE

These three pixel topologies make up a configurable pixel array, and digital logic blocks were designed around this array to create a complete mixed-signal γ photon spectrometry system. The γ photon spectrometer's system architecture is shown in Fig. 5 including the configurable pixel array, $DT$ processing unit, and SRAM sensitivity calibration unit which are described in more detail in the subsequent sections. The system timing diagram is shown in Fig. 6, and the die photo is shown in Fig. 7.

*A. Configurable Pixel Array*

61 rows of the pixel array use the energy-resolving pixel, 10 rows use the low-flux pixel, and 5 rows use the energy-calibrating pixels. Energy-resolving pixels are always enabled to allow for large sensing area, energy-binning of incoming γ photons. Low-flux pixels are enabled to allow for higher count rates when the γ flux in the current environment is less than a global threshold in all the energy-resolving pixels. $DT$ can only be calculated for the low-flux pixels in low-flux environments because there is a larger possibility of two γ detection events in the same pixel at the same time but on different diodes in high-flux environments. The fraction of detection events that are coincident events is given by:

$$\frac{p_{coincident}}{p_{interaction}} = \frac{\sum_{i=2}^{6}\left({}_6C_i \times p_{interaction}^i\right)}{p_{interaction}} = \frac{\sum_{i=2}^{6}\left({}_6C_i \times (1-e^{-\mu_{CS}\Delta x})^i\right)}{1-e^{-\mu_{CS}\Delta x}} \quad (14)$$

where $p_{coincident}$ is the probability of coincident detection events in the low-flux pixels, $p_{interaction}$ is the probability of interaction of a γ photon with a single reverse-biased diode, ${}_6C_i$ is the number of combinations that are formed from picking $i$ diodes from a set of 6, $\mu_{CS}$ is the Compton scattering cross section in silicon, and $\Delta x$ is the depth of the depletion region of the diode. Given a 100 keV (i.e., very high $\mu_{CS}$) photon and a depletion depth of 1.5 μm, the fraction of detection events that are coincident events is 0.002. This means that for every 500 detection events there will be one coincident event in the same pixel. The global threshold to en-

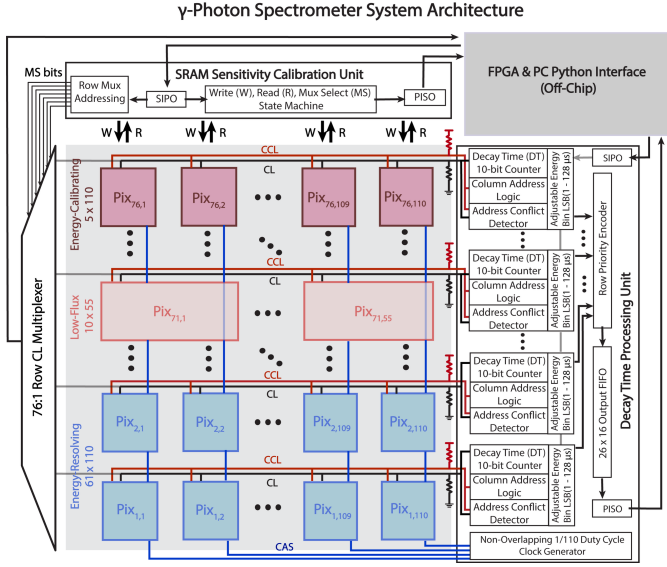

**Fig. 5:** γ Photon Spectrometer System Architecture. System is made up of a configurable pixel array with 3 types of pixels, an SRAM sensitivity calibration unit to set the gain of each in-pixel VGA, and a $DT$ processing unit that handles count data, addressing, $DT$ measurements, and communication off-chip.

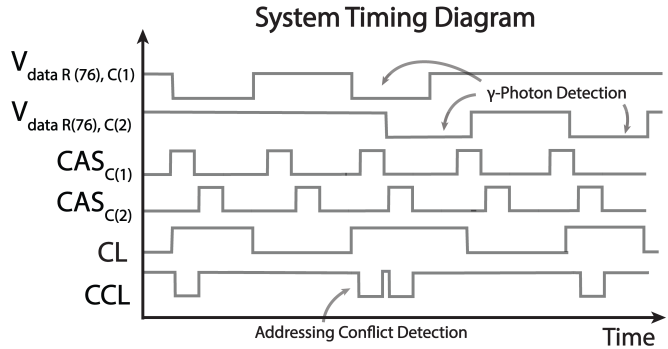

**Fig. 6:** System Timing Diagram. Example of timing diagram for $V_{data}$, CAS, CL, and CCL lines on row (R) 76 and columns (C) 1 and 2 in the pixel array.

-able the low-flux pixels is chosen to ensure that it is unlikely to detect one coincident event in the same pixel given the recording time and incident γ photon flux. This is solely an issue in the low-flux pixels since a coincident event in any of the 6 diodes in the same pixel cannot be resolved. Coincident events affect the resolved energy, so if the γ flux is high enough that a significant number of coincident events are probable, the low-flux pixels will not be enabled.

Pixels in rows 72-76, the energy-calibrating pixels, have 6.25, 5.76, 5.29, 4.84, and 4.41 µm² diodes, respectively. The energy-calibrating pixels are used to calibrate the configurable clock period used for measuring the $DT$ of a γ photon detection event. This is done by measuring the $V_d$ directly using pixel-specific energy-binning (i.e., larger diode sizes can only detect large $E_{dep}$ events) and by measuring the resulting $DT$ using the largest counting clock period initially (i.e., 128 µs). The energy dynamic range of the system is set to be the maximum measured $DT$.

This maximum $DT$ 10-bit value measured during the calibration period, $DT_{cal}$, can be used to find the period that supports the dynamic range of the system using Eqn. 15. $DT_{cal}$ is first scaled to the $C_{diode}$ of an energy-resolving pixel and then converted to µs by multiplying by the maximum period of the configurable counting clock. The period that supports the dynamic range is found by dividing by the dynamic range of the 10-bit counter.

$$T_{cnt,approx} = \frac{DT_{cal} \times \frac{C_{diode}}{C_{diode,en}} \times 128\ \mu s}{2^{10}-1} \quad (15)$$

where $T_{cnt,approx}$ is the approximated ideal clock period to optimize both energy resolution and energy dynamic range, $C_{diode,en}$ is the capacitance of the diode-sensing node of that row, and $C_{diode}$ is the capacitance of the diode-sensing node in the energy-resolving pixels. The closest possible clock period that can be set on-chip is then chosen using Eqn. 16:

$$T_{cnt,optimal} = nn(T_{cnt,approx}, 1-128\ \mu s) \quad (16)$$

where $T_{cnt,optimal}$ is the period of the configurable counting clock used to approximate $DT$ and the $nn$ function is the nearest neighbor function between the calibration result and the available configurable clock periods 1-128 µs.

### B. SRAM Sensitivity Calibration Unit

The SRAM sensitivity calibration unit is used to read from and write to the 3 × 1 in-pixel 6T SRAM in every pixel to control the gain of each VGA in order to maximize the useable sensing area on-chip and reduce acquisition times when rastering the device around the resection site. SRAM read and write commands can be sent to the SRAM control logic serial-input parallel-output (SIPO) block from off-chip, and these bits are responsible for moving the read or write state machine accordingly. There are 3 word-lines (WL) supplied to each row of the pixel array and 2 bit-lines (BL) supplied to each column. Read and write comparators are placed at the base of the pixel array to drive and read the BLs for each column of SRAM bits. The pixel array contains 25,080 bits of SRAM, 110 BL's, and 228 WL's. The SRAM read operation utilizes a parallel-in serial-out (PISO) block to serialize data and send it off-chip.

Each pixel is initially calibrated with no radiation present by enabling one pixel at a time (i.e., setting all pixels 3-bit SRAM to 000 except one) and checking the CL of the corresponding row using a 76×1 row multiplexer that relays any of the 76 CL's off-chip. The output of this multiplexer is connected to an FPGA that detects the number of active high pulses on the CL as one pixel's in-pixel SRAM is incremented. The SRAM value in the one enabled pixel is increased from 0 to 7 to boost the VGA gain until the CL is triggered by noise or starts detecting counts in a 30-second acquisition time. The SRAM is set to one value below this threshold, and this process is repeated for every pixel in the array one at a time. The process is completely automated using the FPGA, and the SRAM is calibrated once. This configuration is saved locally and loaded onto the chip upon power-up. The complete algorithm is summarized below:

---

**Algorithm 1** Pixel Array Sensitivity Calibration
1: Place ASIC in dark environment with no radiation
2: Set in-pixel SRAM for all pixels to 0 (disable)
3: **for** row= $1, 2, \ldots, 76$ **do**
4:     **for** column= $1, 2, \ldots, 110$ **do**
5:        **while** in-pixel SRAM$< 8$ and CCL stays low **do**
6:           Increment SRAM of pixel$_{row, column}$ by 1
7:           Monitor CCL of $row$ for 30 seconds using multiplexer
8:        **end while**
9:        Save SRAM value
10:       Set SRAM of pixel$_{row,column}$ to 0
11:    **end for**
12: **end for**
13: Write saved optimal SRAM values to configurable pixel-array

---

*C. DT Processing Unit*

The $DT$ processing unit finds the row address, column address, and $DT$ of each $\gamma$ photon detection event (Fig. 5). Each row shares a CL and CCL that are routed to this processing unit. In this way, row number encodes the pixel architecture used and that is being read from. The CL outputs a high pulse whenever a $\gamma$ photon has been detected. When CL is high, a 10-bit counter is used to count the number of cycles CL is high by utilizing a configurable period clock. Using a 10-bit counter as opposed to a counter with a smaller bit depth allows for wide dynamic range energy deposition quantification for any of the configurable clock settings. The period of the clock is stored as a one-hot 8-bit number. Using the SIPO block in this processing unit, a 3-bit number can be read in to indicate how many times to bit shift the 8-bit representation of the value 1. In this way, the clock is configurable from 1-128 µs in 2x increments. One-hot configurability allows for adjustable energy resolution but optimizes more for energy-dynamic range and the area of this digital block. This tradeoff was chosen instead of a fully-binary representation because the necessary energy-resolution can be shifted based on the $r_{ds}$ of the PMOS device that reverse-biases each diode in all the pixels, without any increase in chip area. The CCL is an active-low line that is used to discern which column in a specific row a $\gamma$ photon detection event originated from. Column addressing is implemented by generating 1 µs wide, 110 µs period, non-overlapping (i.e., 1µs offset from each other in time), active-high column address sweep (CAS) signals and routing these signals down every column of the pixel array. Whenever a CAS signal is high and a $\gamma$ photon is detected, CCL outputs a low pulse. The CCL and the inverted version of the CAS signal are then compared with an NOR-gate to identify the column a $\gamma$ photon count originated from.

This block also identifies row coincidence detection and column addressing conflicts where there are multiple $\gamma$ photon detection events on the same CL at the same time or the column address is missed, respectively. Row coincidence detection is carried out by identifying times where the CL is high when CCL goes low multiple times. This event implies two coincident $\gamma$ photon detection events in two different pixels in the same row driving the CL high at the same time. The CCL goes low multiple times because the CAS signal goes high in multiple columns when there is a $\gamma$ photon detection even in those columns. Column-addressing misses occur when a CAS signal does not overlap with when $V_{data}$ goes low in the pixel. To guarantee that the column address can be resolved, the decay time has to be at least 110 µs, the period of the non-overlapping column sweeping signal. If the $DT$ is less than 110 µs there is some probability that the column address is missed. Regardless, these conflicts are flagged with a 2-bit header that is sent off-chip: 00 - conflict free, 01 - row coincident event, or 10 - missed column-address.

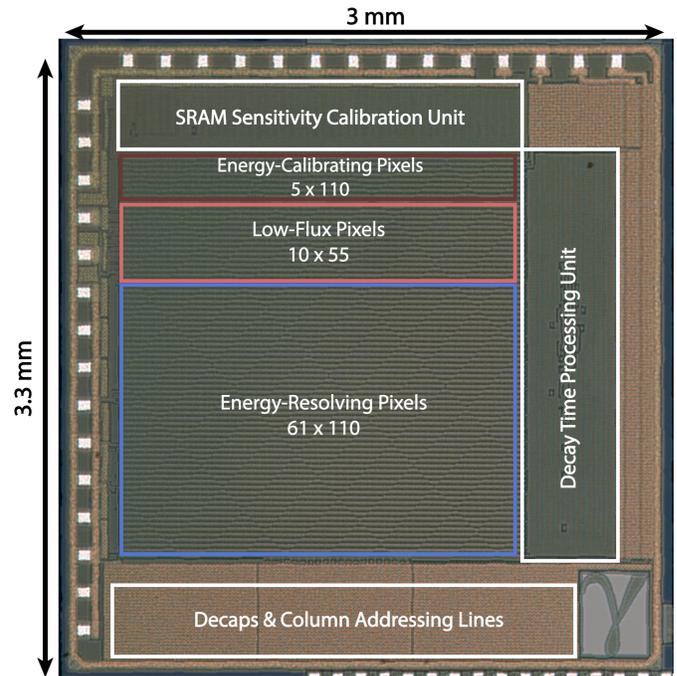

Fig. 7: Die Micrograph of $3 \times 3.3$ mm $\gamma$ Photon Spectrometer. Pixel types, SRAM sensitivity calibration unit, decay time processing unit, and addressing lines are labelled.

All this data is sent to a row priority encoder to prioritize how data is sent off-chip. Lower row values (between 1 and 76) are prioritized over larger row values, and this is because energy-resolving pixels (lowest row numbers) are always enabled and hold all the information about $E_{dep}$ of $\gamma$ detection events, low-flux pixels provide $E_{dep}$ information in low-flux environments but are not always enabled, and energy-calibrating pixels can provide $E_{dep}$ information but are primarily used for system optimization during runtime. Because there is no difference between columns in this pixel array, a column priority encoder is unnecessary and is only based on time-of-arrival. This row priority encoder determines the order in which counts get loaded into a first-in first-out (FIFO) block. The FIFO block has a width of 26 bits (2 bits for conflict flag, 7 bits for row address, 7 bits for column address, 10-bits for $DT$ measurement) and a depth of 16. This prevents new count data from overwriting old count data in high-count environments by providing a buffer between the configurable pixel-array and off-chip FPGA.

## V. SPECTROMETER CHARACTERIZATION WITH $\gamma$-EMITTING RADIOISOTOPES

The pixel architecture was characterized for $SNR_{DT}$ and $\varepsilon_{incident}$ using an on-chip test structure. These measured values were compared to Eqn. 10 and 11, respectively. The spectrometer was then characterized at the UCSF Department of Radiology by using three γ-emitting radioisotopes ($^{64}$Cu, $^{133}$Ba, $^{177}$Lu) with emissions ranging in energy from 31 keV to 1.346 MeV [54]. $^{64}$Cu has two γ photon emissions (511 keV, 1.346 MeV), but its 511 keV emission is dominant. This radioisotope was therefore used to demonstrate the device's linearity with radioactive activity and to characterize the response of the different pixel architectures. $^{133}$Ba has 4 different energies (31 keV, 81 keV, 303 keV, 356 keV) and was used to demonstrate the spectrometer's ability to discern closely spaced incident γ photon energies in this Compton scattering interaction regime. $^{177}$Lu is a common isotope used in cancer radiopharmaceuticals and has two closely spaced energy peaks (113 keV, 208 keV). This radioisotope was used to show the spectrometer's ability to track small changes in the $E_{dep}$ distribution as compared to the $^{133}$Ba source and demonstrate the robustness of the spectrometry method.

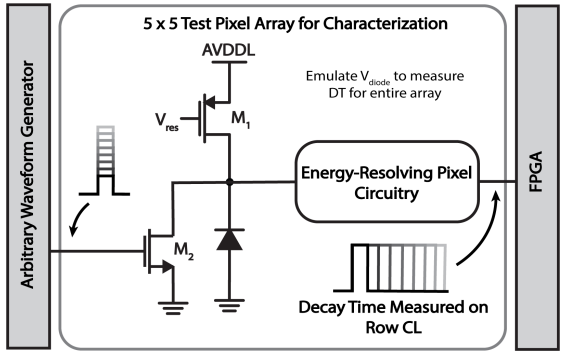

Fig. 8: Pixel Characterization Test-Structure. Function generator and additional NMOS emulate charge generation in the diode depletion region and the resulting $DT$ on each row CL is measured.

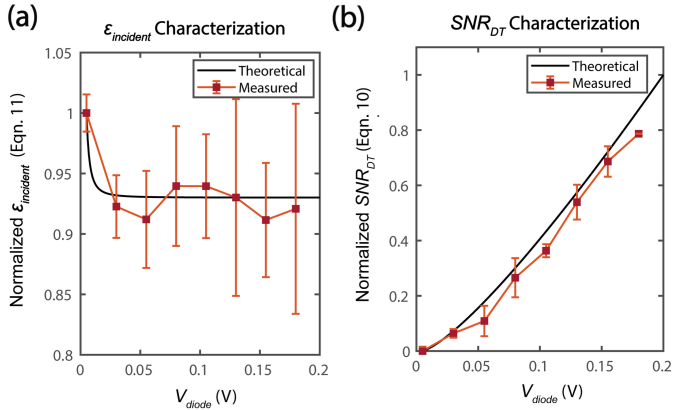

Fig. 9: Pixel Characterization Results. (a) Normalized measured vs. theoretical error in incident energy or $\Delta E_{dep}$. (b) Normalized measured vs. theoretical $SNR_{DT}$.

*A. Pixel Characterization Setup*

A 5 x 5 test pixel array was included on the IC to characterize pixel performance versus the expected $SNR_{DT}$ and $\varepsilon_{incident}$. The pixels in this array were modified with an additional minimum-sized NMOS device (W/L = 220 nm/180 nm) connecting the diode-sensing node to ground to mimic the small current draw (and resulting voltage drop, $V_{diode}$) caused by the interaction of a γ photon with the depletion region of the diode. The gate of this NMOS device is set off-chip using an arbitrary waveform generator (Fig. 8). This modification was not included in the actual pixel array due to the sensitivity drop that would be introduced by the additional parasitic capacitance. $\overline{DT}$ and thereby $\varepsilon_{incident}$ was measured by taking 10 $DT$ measurements on the CL of this 5 x 5 array in response to 10 of the same 1 µs input pulses at the gate of the pull-down NMOS device. This was then repeated for 8 different pulse magnitudes. The $SNR_{DT}$ was calculated at each of these pulse heights by dividing the average $DT$ of each of these measurements by the measured $\overline{DT}$. This was then repeated for all 25 pixels. Each point in Fig. 9(a) and 9(b) is the average $\varepsilon_{incident}$ and $SNR_{DT}$ for a specific input pulse magnitude, respectively, with the pixel-to-pixel variation (i.e., fixed pattern noise) given by the error bars around each point. The theoretical $\varepsilon_{incident}$ and $SNR_{DT}$ are plotted by evaluating the proportionality expressions in Eqn. 11 and 10, respectively.

As $V_{diode}$ increases, the $\varepsilon_{incident}$ decreases slightly (i.e., energy resolution improves) and the $SNR_{DT}$ increases, with the characterization results matching well with the derived expressions. The error between the theoretical and measured $\varepsilon_{incident}$ and $SNR_{DT}$ is likely due to the larger $C_{diode}$ contributed by the additional NMOS capacitance in the measurement results. The error bar sizes on both plots are small due to the SRAM sensitivity calibration technique. Since these test structures have the same in-pixel SRAM and VGA used for sensitivity calibration in the main pixel array, the pixel-to-pixel sensitivity variation across these structures is representative to that of the main pixel array. This calibration prevents the observation of wider extremes of performance or hot pixels that would obscure accurate count statistics.

*B. Experimental γ-Sensing Setup*

Fig. 10 shows a diagram of the experimental setup and Fig. 11 shows the experimental photos of the setup used to test the γ photon spectrometer. The IC-based spectrometer and daughter board are mounted on one end of a motorized x-axis linear stage. A γ-emitting point source of $^{64}$Cu, $^{133}$Ba, or $^{177}$Lu [54] was placed 1 cm away from the spectrometer to emulate a small cluster of cancer cells 1 cm away from the detector. The x-axis linear stage is used to precisely align the spectrometer with the γ-emitting point source at a distance of 0 mm, and is subsequently used to move the source 1 cm away from the IC with sub-mm precision. Pixels were calibrated in the dark and the low-flux pixels were enabled for the whole acquisition time. To test the linearity of the device with radioactive activity or γ photon flux, 7 eppendorf vials with 5 µL of Cu-64 (300, 150, 75, 38, 19, 9, 4.5 µCi) were prepared and measured. The ASIC was then used to continually monitor a vial with 300 µCi of $^{64}$Cu (half-life=12.7 hours) in 5 µL for 107 hours (approximately 8 half-lives). Acquisitions were taken every 5 minutes and compared to $^{64}$Cu's ideal radioactive decay curve. 5-minute acquisitions were used to emulate the acquisition time needed to scan the tumor resection site in the operating room.

Due to its many γ photon emission energies, a 1 µCi $^{133}$Ba point source was used to demonstrate the IC's ability to encode $E_{dep}$ with $DT$. A $^{133}$Ba disk source was placed on the x-axis linear stage and measured for 5 minutes. The $DT$ measurement from all the recorded γ photon detection events are collected as a histogram and converted to a quantity propo-rtional to $E_{dep}$ using Eqn. 7. TOPAS is used to simulate this $^{133}$Ba point source 1 cm away from a silicon detector. The detector is modeled as a 3 mm × 3.3 mm × 500 μm slab of silicon, with γ photons depositing energy in the first 1.5 μm (i.e., depletion region depth) recorded. The $E_{dep}$ distribution using TOPAS is compared to the measured distribution. The simulated TOPAS distribution was then used to assign the proportion of each $E_{dep}$ bin from a specific γ photon energy.

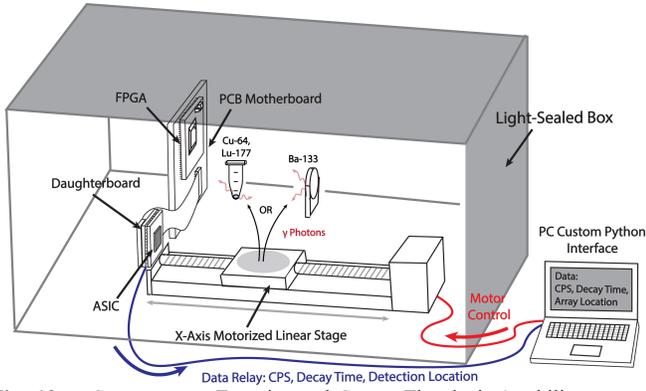

**Fig. 10:** γ Spectrometer Experimental Setup. The device's ability to track radioactive activity and γ photon energies was demonstrated. A point source of $^{64}$Cu, $^{133}$Ba, and $^{177}$Lu was mounted on an X-axis linear stage 1 centimeter away from the spectrometer.

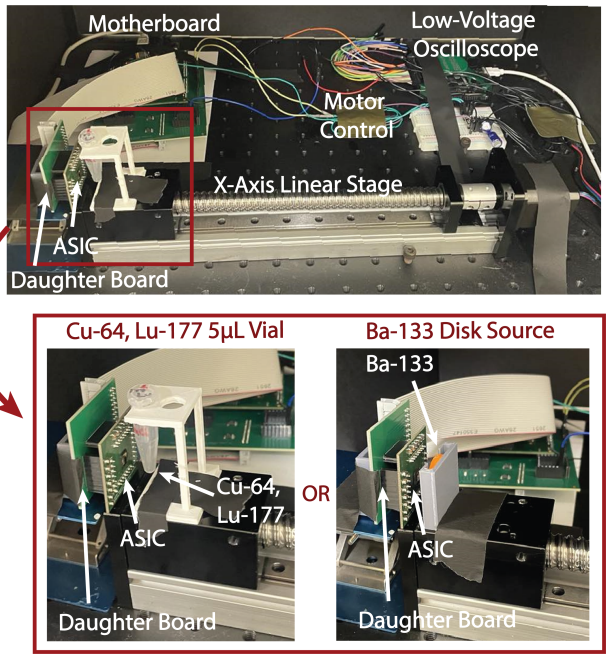

**Fig. 11:** Experimental Photos of Validation Setup. A 5 μL eppendorf vial of $^{64}$Cu or $^{177}$Lu, or a $^{133}$Ba disk source was placed 1 cm from the spectrometer.

Because γ photons are sparsely ionizing, small changes in $E_{dep}$ have to be measured to infer the correct $E_{incident}$. An eppendorf vial containing 1.2 mCi of $^{177}$Lu in 1.2 μL was placed on the x-axis linear stage and measured to show the spectrometer's ability to track the change in the measured $E_{dep}$ distribution compared to $^{133}$Ba, and demonstrate the fidelity of the $E_{dep}$-sensing method. $^{177}$Lu has two γ photon energy emissions in the Compton scattering regime.

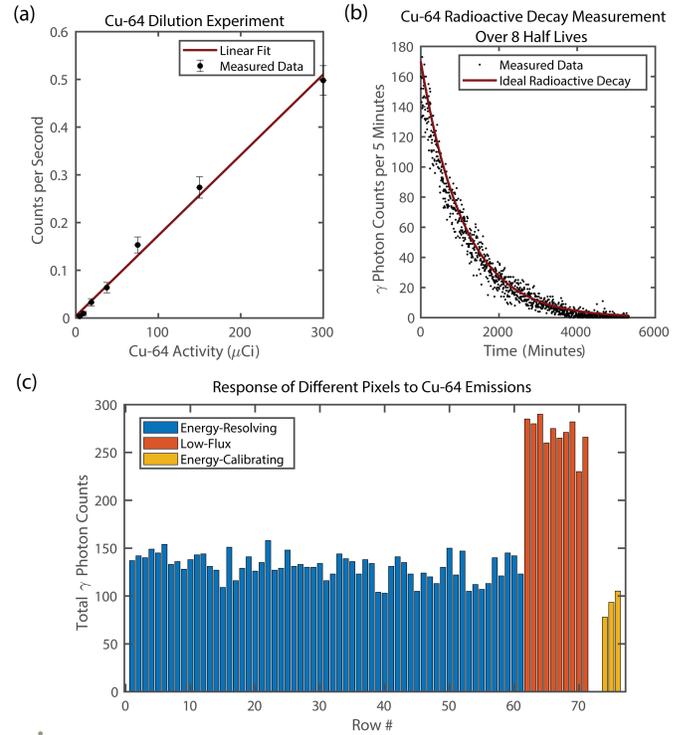

**Fig. 12:** $^{64}$Cu Experimental Results. **(a)** Linearity of γ photon counts per second recorded by spectrometer with $^{64}$Cu activity. **(b)** Ability of spectrometer to track the ideal radioactive decay curve of $^{64}$Cu. **(c)** Total γ photon counts from each row and different pixel architectures in the array.

## C. Experimental γ-Sensing Results

The IC-based spectrometer's counts per second (CPS) were highly linear with the radioactive activity of $^{64}$Cu ($R^2$=0.9942) with CPS varying from 0.008 to 0.5 for activities ranging from 4.5 to 300 μCi, respectively (Fig. 12(a)). The spectrometer accurately tracks the ideal radioactive decay of $^{64}$Cu over 107 hours. The system accurately tracks γ-flux activity levels as low as 1 μCi with 5 minute recordings with +/- 10% error (Fig. 12(b)). The recorded radioactive decay curve is integrated and the response of each pixel row to $^{64}$Cu is reported. Due to the in-pixel SRAM VGA sensitivity calibration, all the energy-resolving pixel rows have similar counts. The low-flux pixel rows have greater than 2x higher counts compared to the energy-resolving pixels given their increased fill-factor but are < 3x higher due to the reduction in SNR at the current summation node (Eqn. 13). The counts in the energy-calibrating pixel rows decrease with the increasing diode sizes, with the 5.76 μm$^2$ and 6.25 μm$^2$ diode rows recording 0 counts over the recording time. The $DT$ from the 5.29 μm$^2$ diode energy-calibrating pixel rows set the period of the configurable clock using Eqn. 15 and 16 by deducing the maximum $E_{dep}$ from any γ emission from $^{64}$Cu. This prevents saturating the 10-bit counter used to measure the $DT$ (Fig. 12(c)).

The γ photon detection events measured 1 cm away from a $^{133}$Ba disk source were collected and the $DT$ measurement for each of these events were used to estimate a quantity proportional to $E_{dep}$, $e^{\frac{DT}{\tau}}$. The $E_{dep}$ in the diode depletion region by the $^{133}$Ba γ photon emissions were simulated using TOPAS by modeling a point source 1 cm away that emits 4 γ photon energies with the same probability of emission as seen experimentally: 31 keV (13.8%), 81 keV (32.9%), 302 keV (18.3%), 356 keV (62.1%). The simulated $E_{dep}$ with sub-keV energy bins ($\Delta E_{bin}$ = 0.1 keV, $\Delta E_{dep} \approx 0.2$ keV) was very proportional to the measured $e^{\frac{DT}{\tau}}$ distribution (Fig. 13(a)). Both distributions have a low and higher energy $E_{dep}$ peak, with their tails being nearly identical. The finite mismatch between the distributions for some of the $E_{dep}$ bins between 0 and 2 keV is most likely due to the fact that the SiO$_2$ and back end of line (BEOL) metal interconnects in the silicon are not simulated which can cause additional scattering and attenuation of incoming γ photons that may eliminate low-energy $E_{dep}$ events in the simulated distribution, and due to the quantization error in sensitivity introduced by the in-pixel DACs. Because simulated $E_{dep}$ can be labeled by the incident γ energy that deposited it (Fig. 13(b)), the simulation can be used to identify the incident γ energy spectrum in the measured $e^{\frac{DT}{\tau}}$ distribution (Fig. 13(c)).

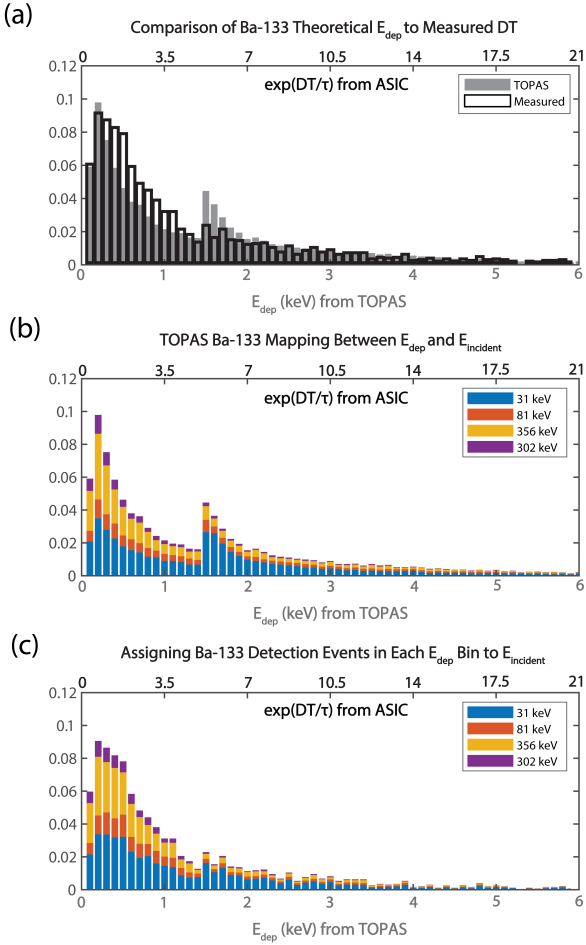

**Fig. 13:** $^{133}$Ba Experimental Results. **(a)** Similarity between DT-derived $E_{dep}$ histogram and $E_{dep}$ histogram from TOPAS for $^{133}$Ba. **(b)** TOPAS-derived mapping between $E_{dep}$ and $E_{incident}$ for $^{133}$Ba. **(c)** Measured $E_{incident}$ histogram using TOPAS mapping for $^{133}$Ba.

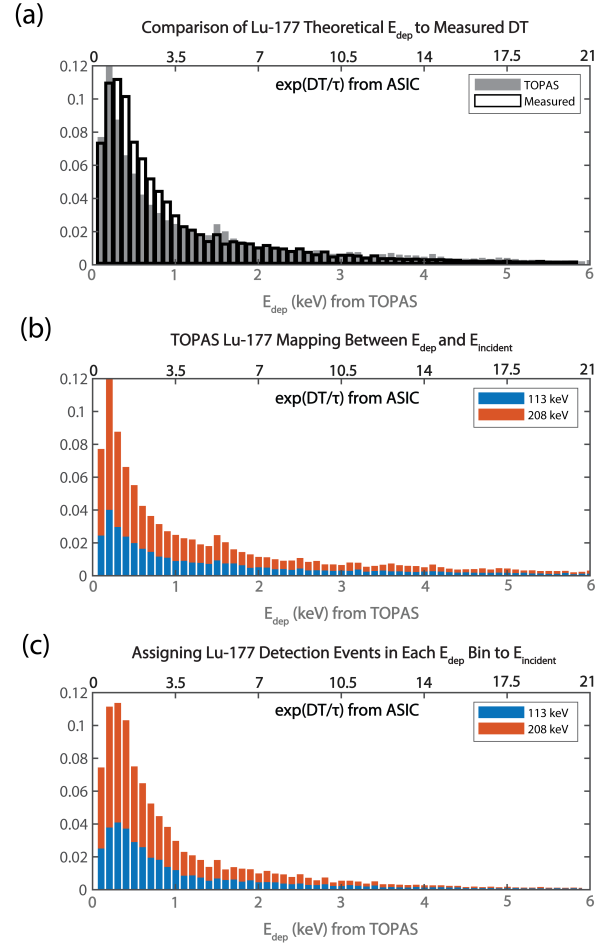

**Fig. 14:** $^{177}$Lu Experimental Results. **(a)** Similarity between DT-derived $E_{dep}$ histogram and $E_{dep}$ histogram from TOPAS for $^{177}$Lu. **(b)** TOPAS-derived mapping between $E_{dep}$ and $E_{incident}$ for $^{177}$Lu. **(c)** Measured $E_{incident}$ histogram using TOPAS mapping for $^{177}$Lu.

Similarly, the γ photon detection events measured 1 cm away from a $^{177}$Lu vial were collected and the $DT$ measurement for each of these events were used to estimate $e^{\frac{DT}{\tau}}$. A similar TOPAS simulation with the γ photon emissions from a $^{177}$Lu point source with the same probabilities of emission as seen experimentally was run: 113 keV (6.6%), 208 keV (11.1%). The spectrometer is able to accurately capture the small changes in $E_{dep}$ when measuring $^{177}$Lu compared to $^{133}$Ba, with a larger proportion of γ detection events from $^{177}$Lu depositing lower amounts of energy (Fig. 14(a)). The simulation can be used to identify the incident γ energy spectrum (Fig. 14(b)) in the measured $e^{\frac{DT}{\tau}}$ distribution (Fig. 14(c)).

The sensitivity of the developed spectrometer is comparable to larger γ photon counters used in clinics for sentinel lymph node (SLN) monitoring. Many γ photon counters for SLN drainage monitoring require a minimum sensitivity of 1.4 μCi at a depth of 5 mm [27][55]. The lowest activity tested and that was shown to be resolvable in this paper was 1 μCi at 1 cm away with a 5-minute acquisition time. The energy resolution achieved with this system is also compatible with the necessary depth resolution needed for clinical RGS [56]. The $E_{incident}$ resolution is given by the minimum change in the $E_{dep}$ distribution that can be sensed. This is therefore given by the $E_{dep}$ resolution (Eqn. 11, Fig. 8), which has been shown to be sub-keV (i.e., approximately 0.2 keV) from the measured distributions with $^{133}$Ba (Fig. 13) and $^{177}$Lu (Fig. 14). This energy resolution corresponds to a depth resolution of less than 1 mm for γ photon energies between 100 keV and 1 MeV (i.e., Compton scattering regime). This aligns well with the maximum precision (i.e., 0.8 mm [57]) of many surgical tools used for tumor resection.

TABLE I: COMPARISON OF STATE-OF-THE-ART γ PHOTON SPECTROMETERS

| Performance Metric | [27] | [28] | [45] | [47] | This Work |
|---|---|---|---|---|---|
| Application | Radioguided Sentinel Node Resection | Radioguided Cancer Resection | X-Ray External Beam Radiotherapy | Digital X-Ray Imaging | Radioguided Cancer Resection |
| Flux-Level | Low (μCi) -Equivalent to nGy/min | Low (μCi) -Equivalent to nGy/min | High (cGy/min) | High (μGy/min) | Low (μCi) -Equivalent to nGy/min |
| Detection Method | APD + Scintillator | SiPM + Scintillator | Diode + VDD Grid | Photoconductor | Diode with DT Quanification |
| Max Fill Factor | Single Diode | Photomultiplier | 2% | Photoconductor | 2.5% |
| Total Number of Pixels | 1 | 1 | 4180 | 16384 | 7810 |
| Number of Energy Bins | No Energy Resolution | No Energy Resolution | 4180 | 3 | Up to 8184 DT Bins 5 Coarse Bins |
| Pixels per Energy Bin | No Energy Resolution | No Energy Resolution | 1 | 16384 | 7260 DT Pixels 550 Coarse Pixels |
| Energy Range of Operation | 122.5-157.5 keV | 2.28 MeV Beta Particle | 50-100 keV | ~49 keV | 31-1346 keV |
| Configurable Energy Dynamic Range | No | No | No | No | Yes (1μs-130ms) |
| Configurable Energy Resolution | No | No | No | No | Yes (1-128μs) |
| Single γ Photon Sensitive | No | No (Beta Particle Sensitive) | Single X-ray Sensitive | No | Yes |
| Power Dissipation | Not Specified | Not Specified | 0.822 mW | 75 mW | 0.6 mW |
| Direct Detection in Si | No | No | Yes | No (Uses $HgI_2$) | Yes |
| Form Factor | 12 x 12 x 40 mm | 12 x 12 x 150 mm | 3 x 3 mm | 8.8 x 8.8 mm | 3 x 3.3 mm |

## VI. CONCLUSIONS

A 3×3.3 mm single γ photon sensitive spectrometer was designed and fabricated in 180 nm CMOS technology to enable precision, real-time detection and removal of residual cancer cells tagged with a γ-emitting radioisotope during RGS. The system integrates (1) a configurable analog pixel array that optimizes the system based on the γ photon flux, and the necessary energy resolution and energy dynamic range, (2) an SRAM sensitivity calibration unit which can read and write to in-pixel 6T-SRAM that sets the gain of each pixel's VGA individually to set the sensitivity of the whole pixel array to be approximately the same, and (3) a DT processing unit to measure the DT of each γ photon detection event and relay this data off-chip. This mm-scale γ photon spectrometer consumes an average static power of 0.6 mW with the chip calibrated for runtime (510 μW for the configurable analog pixel array, 20 μW for the SRAM sensitivity calibration unit, and 70 μW for the DT processing unit), on an analog and digital VDD of 1.2V. Power was optimized in the configurable pixel array by increasing sensitivity of the diode-sensing node such that low in-pixel DAC positions are utilized more often and hence small static-current is consumed per pixel. The DT-based spectrometry method helps avoid in-pixel ADCs that would consume much more power than the inverters and digital counters implemented in-pixel and in the DT processing unit. The SRAM sensitivity calibration unit is placed in an 'idle' state when not reading or writing from the SRAM and therefore consumes very little power during runtime.

A comparison table with state-of-art radiation detectors and dosimeters is summarized in Table I, and includes radioguided surgery ionizing radiation detectors [27][28] for application-specific comparison and IC-based energy-binning ionizing radiation detectors [45][47] for system performance comparison. This paper presents the first mm-scale, RGS-compliant γ photon spectrometer with the widest energy detection range and energy-binning area, lowest total static power consumption, and is the only system with configurable sensitivity, sensing area, energy resolution, and energy dynamic range depending on the radioisotope used and specific surgical scenarios. The spectrometer was shown to be able to resolve $DT$ with high SNR, calculating the $E_{dep}$ distribution in the depletion region with sub-keV accuracy. The spectrometer's ability to configure its settings to optimize γ photon-sensing performance, to detect single γ photons, resolve their $E_{dep}$ in the diode depletion region, and map this $E_{dep}$ back to $E_{incident}$ was demonstrated with low activities (~ 1 μCi) of three γ-emitting radioisotopes, $^{64}$Cu, $^{133}$Ba, and $^{177}$Lu, with emissions in a wide energy range from 31 keV to 1.346 MeV. The device is applicable to the detection and spectrometry of any γ-emitting radioisotope.

The developed γ photon spectrometer achieves (1) a large, high-sensitivity sensing area (i.e., 8360 sensing elements in pixel array), (2) small form factor (i.e., 3×3.3 mm) and low power (i.e., 0.6 mW), (3) high energy resolution (i.e., sub-keV accuracy in $E_{dep}$) (4) large energy dynamic range (i.e., $DT$ logarithmic compression), and (5) configurability in sensitivity (i.e., in-pixel VGA sensitivity; low-flux pixels), energy resolution, and energy dynamic range (i.e., energy calibrating pixels and configurable period $DT$ counting clock). It is envisioned that this spectrometer will enable RGS workflows that ensure precision resection of the primary tumor, remnant cancer cells in the margin, and any microinvasion of cancer cells multiple centimeters below the tissue surface. Precision resection of cancer cells would prevent excessive removal of healthy tissues and reduce the need for additional treatment intervention post-surgery.

ACKNOWLEDGMENT

The authors appreciate technical discussions and advice from Dr. Kyoungtae Lee and Dr. Apurva Pandey. The authors would like to thank the Berkeley Wireless Research Center (BWRC), Berkeley Sensors and Actuators Center (BSAC), and the University of California, San Francisco (UCSF) Department of Radiology and Biomedical Engineering and Department of Radiation Oncology for providing resources and infrastructure that helped conduct this study.

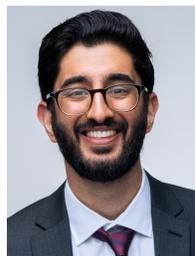

**Rahul Lall** (Member, IEEE) received his B.S.H. degree in Electrical Engineering from Stanford University in 2021, graduating as a Frederick E. Terman Scholar and ranking in the top 5% of his graduating class. He received his M.S. and Ph.D. in Electrical Engineering and Computer Sciences from the University of California, Berkeley in 2023 and 2025, respectively. His research focuses on developing biomedical circuits and sensors for optimizing cancer radiotherapy. Rahul has previously worked on carbon nanotube transistor reliability (Stanford, 2018-19), flexible electronics (GE Research Center, 2019), and heterogenous integration (IBM AI Hardware Research Center, 2020-21). He was the recipient of the NSF Graduate Research Fellowship in 2021, the recipient of the IEEE Intersociety Conference on Thermal and Thermomechanical Phenomena in Electronic Systems Best Paper Award in 2022, selected Best of Physics at the American Society for Radiation Oncology Annual Meeting in 2023, the IEEE Solid-State Circuits Society Predoctoral Achievement Award, and the Best Student Paper

Award at the 2024 Symposium on VLSI Technology and Circuits. He is now a NIH/NCI T32 Molecular Imaging Postdoctoral Fellow at Stanford University with a focus on magnetic approaches to cancer molecular imaging.

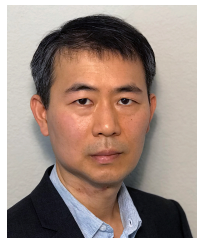

**Youngho Seo** (Senior Member, IEEE), PhD, is a Professor and Director of Nuclear Imaging Physics in the Department of Radiology and Biomedical Imaging at UCSF, and Professor in the Department of Nuclear Engineering at UC, Berkeley. He received his bachelor's degree in Physics from KAIST in South Korea, and completed his PhD in Physics with the dissertation on a dark matter experiment using dual-phase xenon from UCLA. He joined UCSF Physics Research Laboratory in 2003 as a postdoc before joining the faculty in 2006. He leads a group of physicists and engineers working in the field of radionuclide and x-ray imaging instrumentation and physics, and directs the UCSF PRL. His primary research focus is to use quantitative SPECT/CT, PET/CT, and PET/MR molecular imaging tools for a broad range of research areas from small animal imaging using dedicated animal imaging systems and basic instrumentation development for physics analysis of clinical research data.

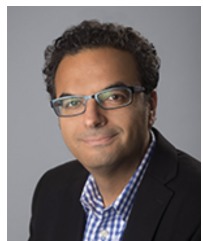

**Ali M. Niknejad** (Fellow, IEEE) received the B.S.E.E. degree from the University of California at Los Angeles, Los Angeles, CA, USA, in 1994, and the master's and Ph.D. degrees in electrical engineering from the University of California at Berkeley (UC Berkeley), Berkeley, CA, in 1997 and 2000, respectively. He is currently a Professor with the EECS Department, UC Berkeley, the Faculty Director of the Berkeley Wireless Research Center (BWRC), Berkeley, and the Associate Director of the Center for Ubiquitous Connectivity. His research interests include wireless and broadband communications and biomedical imaging and sensors, integrated circuit technology (analog, RF, mixed signal, and mm-wave), device physics and compact modeling, and applied electromagnetics. Prof. Niknejad and his coauthors received the 2017 IEEE Transactions on Circuits and Systems—I: Regular Papers Darlington Best Paper Award, the 2017 Most Frequently Cited Paper Award (2010–2016) at the Symposium on VLSI Circuits, and the CICC 2015 Best Invited Paper Award. He was a recipient of the 2012 ASEE Frederick Emmons Terman Award for his textbook on electromagnetics and RF integrated circuits. He was a co-recipient of the 2013 Jack Kilby Award for Outstanding Student Paper for his work on an efficient Quadrature Digital Spatial Modulator at 60 GHz, the 2010 Jack Kilby Award for Outstanding Student Paper for his work on a 90-GHz pulser with 30 GHz of bandwidth for medical imaging, and the Outstanding Technology Directions Paper at ISSCC 2004 for co-developing a modeling approach for devices up to 65 GHz.

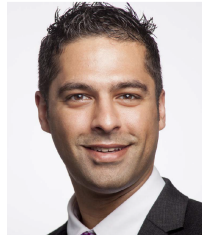

**Mekhail Anwar** (Member, IEEE) received the B.A. degree in physics from the University of California Berkeley (UC Berkeley), Berkeley, CA, USA, where he graduated as the University Medalist, the Ph.D. degree in electrical engineering and computer sciences from the Massachusetts Institute of Technology, Cambridge, MA, USA, in 2007, and the M.D. degree from the University of California San Francisco (UCSF), San Francisco, CA, in 2009. In 2014, he completed a Radiation Oncology residency with UCSF. In 2014, he joined the faculty with the Department of Radiation Oncology, UCSF, with a joint appointment in Electrical Engineering and Computer Sciences at UC Berkeley (in 2021), where he is currently an Associate Professor. His research focuses on developing sensors to guide cancer care using integrated-circuit based platforms. His research centers on directing precision cancer therapy using integrated circuit-based platforms to guide therapy. His work in chip scale imaging has been recognized with awards from the DOD (Physician Research Award) and the NIH (Trailblazer), and in 2020 he was awarded the prestigious DP2 New Innovator Award for work on implantable imagers. He is board certified in Radiation Oncology and maintains a clinical practice specializing in the treatment of GI malignancies with precision radiotherapy.